# Electro-Mechanical Response of a 3D Nerve Bundle Model to Mechanical Loads Leading to Axonal Injury

I. Cinelli, M. Destrade, M. Duffy, and P. McHugh

*Abstract*— *Objective:* Traumatic brain injuries and damage are major causes of death and disability. We propose a 3D fully coupled electro-mechanical model of a nerve bundle to investigate the electrophysiological impairments due to trauma at the cellular level. *Methods:* The coupling is based on a thermal analogy of the neural electrical activity by using the finite element software Abaqus CAE 6.13-3. The model includes a real-time coupling, modulated threshold for spiking activation and independent alteration of the electrical properties for each 3-layer fibre within a nerve bundle as a function of strain. *Results:* Results of the coupled electro-mechanical model are validated with previously published experimental results of damaged axons. The cases of compression and tension are simulated here, to induce (mild, moderate and severe) damage at the nerve membrane of a nerve bundle, made of four fibres. Changes in strain, stress distribution, and neural activity are investigated for myelinated and unmyelinated nerve fibres, by considering the cases of an intact and of a traumatized nerve membrane. *Conclusion:* A fully coupled electro-mechanical modelling approach is established to provide insights into crucial aspects of neural activity at the cellular level due to traumatic brain injury. *Significance:* One of the key findings is the 3D distribution of residual stresses and strains at the membrane of each fibre due to mechanically-induced electrophysiological impairments, and its impact on signal transmission.

## I. INTRODUCTION

Traumatic brain injury (TBI) is caused by mechanical loading to the head due to a sudden acceleration or a blast wave, for example, causing pathologies which range from focal damage of brain tissue to widespread axonal injury [1], [2]. TBI in humans may result from falls, vehicle accidents, sport injuries, military incidents, etc. [1], [3], [4].

The diffuse form of TBI is called Diffuse Axonal Injury (DAI), i.e. a mechanical pathogenesis of an axonal injury, initiated in the deep white matter regions of the brain at the moment of injury [1], [3]. DAI is the most common pathological feature of the mild and severe cases of TBI at cellular and subcellular levels [1], [5], with progressive course [3], responsible for long-lasting neurological impairments following TBI, and high rates of mortality [5]–[7]. Pathological features of DAI include a wide-ranging variety of tissue lesions of the white matter (such as swelling of axonal mitochondria, thinning of the intermodal axoplasm, and development of myelin intrusions [8]), focal haemorrhages, contusions and other brain injuries [1], [3], [7].

Although DAI is classified as an independent category of disease from TBI, the pathological mechanism of DAI is very complex and, currently, there are no diagnostic criteria [3]. A better understanding of

All the authors are with National University of Ireland, Galway (Rep. of Ireland), (email : i.cinelli1@nuigalway.ie; michel.destrade@nuigalway.ie; maeve.duffy@nuigalway.ie ; peter.mchugh@nuigalway.ie; )

mechanically-induced electrophysiological impairments and damage associated with morphological changes of neural cells is urgently needed to improve diagnosis, clinical treatments and prognosis [3], [7].

A number of experimental models of axonal injury reproduce DAI by experimental traumatic insult, where a Traumatic Axonal Injury (TAI) induces DAI to investigate the relationship between mechanical forces and structural and functional responses of axons in experiments [7]. Experimental studies conducted on single axons [9] and nerve fibres [10] aim to induce TAI by applying pressure [11], [12], displacement [1], [10], strain [9], [13], shear strain [14] and electroporation [15]. Although different types of loads seem to initiate TAI, recent studies have shown that the degree of neuronal impairment is directly related to the magnitude and rate of axonal stretch [1], [9], [10], [16]. Beyond a critical threshold [10], axons appear to be vulnerable to stretch-induced changes that induce morphological changes at the microscale level, increasing axolemma permeability [1], [3], [5].

The alteration of axolemma permeability is evidence of mitochondrial damage and neurofilament compaction [7], [8]. Axonal damage is a common manifestation of DAI. Injury-induced axonal damage involves damage of the axonal cytoskeleton, resulting in a loss of membrane integrity and impairment of axoplasmic transport, leading to changes in electrical signal propagation [4], [9], [17]. The alteration of neural activity in a mechanically-injured nerve is called neurotrauma [9], [17], [18]. Although recent experimental studies highlight complex electro-mechanical phenomena at the nerve membrane layer [19], [20], the injury-induced electrophysiological changes of the electro-mechanical activity in neural cells are poorly understood [7], [9]. Quantifying the induced electro-mechanical changes can help to differentiate severity of injury and to understand the alteration in signal propagation at cellular and nerve bundle levels.

Computational electro-mechanical models, coupling mechanics and electrophysiology, are powerful tools to investigate and evaluate neurophysiological, neuropathological processes and neurocognitive damage associated with DAI due to injury at the macroscale [1], [2]. This work presents a novel approach for evaluating and quantifying the changes in neural activity due to TAI by using finite element (FE) models. Previous modeling efforts have simulated one-dimensional damage of a nerve fibre [2] and two-dimensional axonal injury of brain tissue [1]. In this work, using the FE software Abaqus CAE 6.13-3, advanced three-dimensional (3D) models explain the link between TBI and DAI at the microscale level [1], [2], by considering the case of TAI at the axonal and bundle levels. Our 3D FE model of a nerve bundle includes a representation of nervous cells made of extracellular media (ECM), membrane, and intracellular media (ICM). As in [21], these components have finite thicknesses, so that the resulting model is a 3-layer nerve bundle. The bundle model here is a section of an idealized geometry of a nerve bundle, which consists of four parallel cylindrical unmyelinated or myelinated fibres, see Fig. 1. In this paper, a series of mechanical loads (such as pressure [11] and displacement [10]) are applied to the bundle to induce a certain level of damage at the nerve membrane of a fibre, altering the fibre activation dynamics and transmission [16], [17].

This model presents a unique framework for investigating how changes in strain and stress distributions alter the function of 3D myelinated and unmyelinated nerve fibres and bundles. In contrast to previous studies [1], [2], this model includes full coupling between mechanical and electrical domains [22], [23] where the neural electrical activity, as described in [24], is coupled to the mechanical domain through piezoelectricity [20] and electrostriction [19], reproducing biophysical phenomena accompanying the action potentials as observed experimentally [20], [25]. Here, the electrical activity is directly coupled to mechanical deformation by using electro-thermal equivalences in a coupled thermo-mechanical FE analysis [26]. Electro-thermal equivalences and equivalent material properties have been shown to provide an efficient approach to resolve 3D electrical problems in a coupled electro-mechanical analysis in Abaqus CAE 6.13-3 [26].

This paper builds on the work reported in [22], [23]. Here, we report details of how electro-mechanical coupling is represented through thermo-mechanical equivalence; the nerve bundle model and its geometry; and the damage criterion and its implementation.

The electro-mechanical coupling approach is validated based on the strain-based damage criterion [2], [16], [22]. This criterion refers to traumatic mechanically-induced damage on a nerve fibre (refer to [2], [16]) in which the resting ionic potentials of the Hodgkin and Huxley (HH) model are shifted by $20\ mV$, simulating experimental evidence of damaged nerve fibres [16], [17], as in [2]. The injury threshold takes into account axonal strain along the nerve fibre length only [1], [2]. The strain along the fibre length has been shown to be a physiologically relevant injury criterion for multiscale TBI models [1], [2]. Two cases, (i) dynamic pressure and (ii) displacement loads at the bundle level, in which only one fibre is activated, are considered.

This approach has the potential to generate useful insights in studying the mechanics behind neurophysiology, as observed experimentally in damaged nerve membranes of clinical cases (such as multiple sclerosis) [9], [18], [27].

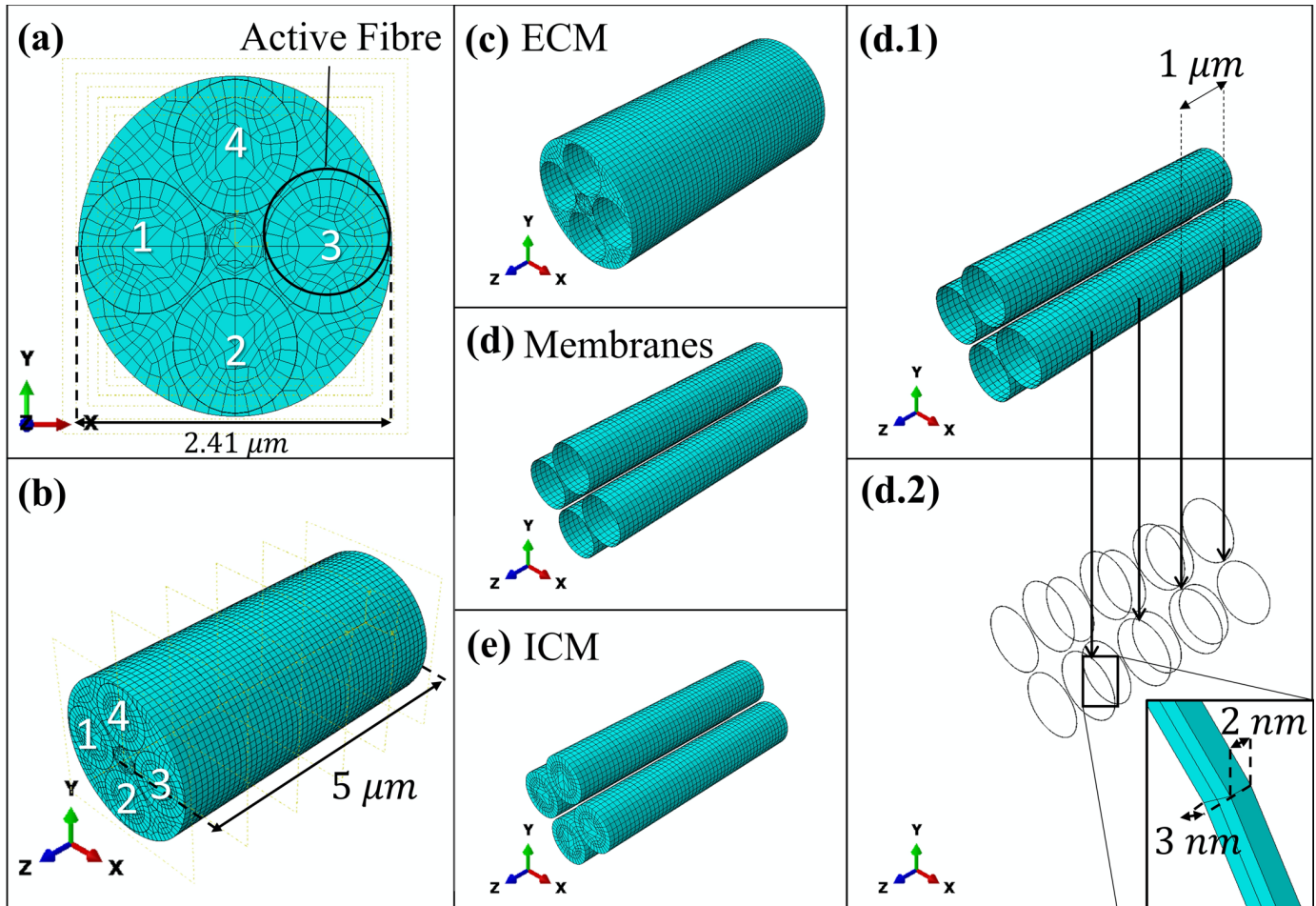

Fig. 1. (a): frontal view and (b): isometric view of the 3-layer nerve bundle, made of four fibres. Fibre #3 is the active fibre, i.e. the fibre activated by a Gaussian voltage distribution [25]. Fibres #1, #2 and #4 are activated by the charges diffusing from Fibre #3. (c): the ECM; (d): the membrane; (e): the ICM. In the case of myelinated fibres, the membrane layer is periodically-partitioned along the fibre length to model the insulation sheath of the myelin layer, see (d.1), and the Ranvier node, see (d.2). The myelin layer length is 1 $\mu m$ and the Ranvier's node length is 2 $nm$, while the radial thickness of the layer is equal to 3 $nm$ [26], [28].

## II. METHOD

*A. Electro-mechanical Coupling*

The electro-mechanical coupling of the action potential [24] is achieved through modelling of the nerve membrane as a piezoelectric material [19], building on the equivalent electro-thermal modelling approach for predicting the electrical behaviour of nervous cells described in [21], [22], [26].

This can be described in terms of the piezo-elastic relation given in equation (1) where $\varepsilon$ is the total strain vector, $\beta$ is the compliance matrix, $\sigma$ is the mechanical stress vector, $\delta$ is the piezoelectric strain coefficient vector, $h$ is the thickness of the piezoelectric layer, and $\Delta V$ is the voltage difference across it [29], [30].

$$\varepsilon = \beta\sigma + \delta(\Delta V/h) \quad (1)$$

Similarly, the thermo-elastic strain-stress relation is described in (2), where $\boldsymbol{\alpha}$ is the thermal expansion coefficients vector and $\mathit{\Delta T}$ is the temperature difference [29], [30].

$$\boldsymbol{\varepsilon} = \boldsymbol{\beta\sigma} + \boldsymbol{\alpha}\Delta T \qquad (2)$$

By the electro-thermal analogy [22], [26], [29], [30], the electric field, approximated here as the voltage across the membrane divided by its thickness, is equivalent to a thermal load, while the piezoelectric constants are equivalent to the thermal expansion coefficients, see (1) and (2).

Simulating the nerve membrane as the dielectric component of a parallel plate capacitor [24], [29], the displacement follows the gradient of the electric field using the approach presented in [29], [30], see (2). That is, the piezoelectric effect is only relevant in the through-thickness direction, represented here with orthotropic piezoelectric constants, approximately $1\ nm$ per $100\ mV$ [20] in the thickness direction and zero in the longitudinal and circumferential directions. Meanwhile, incompressible isotropic mechanical properties [25] are assumed.

As the electrical depolarization is initiated in the nervous cell [24], compressive forces on the nerve membrane arise from piezoelectricity [20], see (2), leading to changes in shape (electrostriction [20], [25], [31]) and corresponding changes in capacitance. The electrical capacitance per unit area, $C$, changes as the square of the voltage [25], [31], see (3) and Fig. 2. Here, $\vartheta$ is the fractional increase in capacitance per square volt (equal to $0.036\ V^{-2}$ [31]) and $\Delta\varphi$ is the surface potential difference (about $-70\ mV$ [31]).

$$C(V) = C(0)[1 + \vartheta(V + \Delta\varphi)^2] \qquad (3)$$

Hence, this approach establishes a fully electro-mechanical coupling accounting, at the same time, for the effect of mechanical loads on electrical response and the effect of the electrical activity on the nerve structure.

*B. Model*

The bundle model simulates the exchange of charges in four identical fibres as shown in Fig. 1. Each fibre consists of a cylindrical region of ICM enclosed by a thin membrane and surrounded by a region of ECM [26].

Two bundle models are used in this study: a fully unmyelinated bundle and a fully myelinated bundle. The neurite radii are: $a_{ICM} = 0.477\ \mu m$, $a_M = 0.480\ \mu m$ and $a_{ECM} = 0.500\ \mu m$ [26] for the ICM, membrane and ECM layers, respectively. As a first step, this analysis is focused on the radial distribution of charges rather than on longitudinal variations. The length of the bundle is $5\ \mu m$ for a diameter equal to $2.41\ \mu m$ in both cases, see Fig. 1, within the range of the human optic axon [32].

To model a myelinated fibre, we assume an ICM enclosed by a single layer, which is periodically-partitioned along the fibre length, similarly to the histologic section of a myelinated fibre, see Fig. 1 (d.1) and (d.2). The insulation sheath of myelin around the nerve fibre is modelled as an insulating layer, which replaces the membrane layer at regular intervals along the fibre [28], see Fig. 1 (d.1). Different conductivity values are assigned to denote the myelin and membrane sections [28], see Fig. 1 (d.2). The width of the piecewise conductive membrane regions (or Ranvier's nodes) is $0.002\ \mu m$ and the internode distance is $1\ \mu m$ [28], see Fig. 1 (d.2).

Incompressible isotropic mechanical properties [25] are assumed in both models, with Young's Modulus equal to 1GPa and a Poisson's ratio of 0.49 [25] (close to incompressibility). Then, the electrical model parameters for unmyelinated and myelinated fibres are taken from [33] and [2], respectively. The resistance per unit length of ICM and ECM is $1.89\ \Omega m$, and for the myelin it is $4\ M\Omega m$. Then, the capacitance per unit area is $0.148\ \mu F/cm^2$ for the ICM, $3.54\ \mu F/cm^2$ for the ECM, and $3.58\ 10^{-3}\ \mu F/cm^2$ for the myelin layer. The corresponding value of the membrane are the HH model values [24] for the steady state and are dependent on voltage, strain and spatial coordinates during loading conditions, see section D.

According to previous published work [34], [35], a neurite refers to a part of the body of a nervous cell (axon) that consists of two structurally distinct regions (axoplasm and membrane). In contrast, a *fibre* usually refers to an axon (as axoplasm and membrane) with a myelin layer. In this work, neurite refers to myelinated fibre too, because the myelin layer is modelled as periodical insulating layers at the nerve membrane layer, whose radial extension has been neglected, as in [28].

*C. Damage Evaluation*

Strain-based damage affects sodium and potassium reversal potentials ($E_{Na}(\varepsilon_m)$ and $E_K(\varepsilon_m)$), see (4), of the HH model, simulating the changes in ionic concentration across the nerve membrane depending on the membrane strain ($\varepsilon_m$) [2]. The resting potentials at physiological conditions are $E_{Na0}$ and $E_{K0}$ [2], [24] see (4). If a maximum strain, $\tilde{\varepsilon}$, is exceeded, then the reversal potentials are zero, otherwise the changes follow the equation (4) if $\varepsilon_m < \tilde{\varepsilon}$ [2]:

$$\begin{cases} E_{Na}(\varepsilon_m) = E_{Na0}(1 - (\varepsilon_m/\tilde{\varepsilon})^\gamma) \\ E_K(\varepsilon_m) = E_{K0}(1 - (\varepsilon_m/\tilde{\varepsilon})^\gamma) \end{cases} \quad (4)$$

Here, the strain threshold, $\tilde{\varepsilon}$, is set at 0.21 as an indicator of the onset of functional damage at which there is a 25% probability of having morphological injury during an uniaxial displacement test of a nerve fibre [10]. The parameter $\gamma$, equal to 2 [2], is an index referring to the sensitivity of the damage to small versus large deformation, see [2]. Additionally, the reversal potential of the leak ions $E_l$- is not influenced by the strain but varies based on changes in gradient concentrations of potassium and sodium across the membrane [2], see Fig. 2. The changes for the leak ions can be derived from the sum of the total membrane currents at resting conditions [2], [24].

*D. Implementation*

The implementation of the coupled Hodgkin and Huxley (HH) model is shown in Fig. 2 (on the right), in contrast to the uncoupled HH model (shown on the left). By using the electro-thermal equivalence [22], [26], the implementation of the neural activity, the distribution of voltage, and the generated strain can be seen in 3D by using well established coupled thermo-mechanical software simulation tools. In the coupled model [22], the membrane electrical conductivity changes in response to the membrane voltage produced during the action potential, incorporating the effects of strain [2], [24], according to (4), while the electrical capacitance per unit area changes with the square of the voltage [31], see (3). The HH reversal voltage potentials change due to the strain at the membrane [2], [16], hence the threshold of spike initiation changes as in [24]. In all cases, strain is caused by membrane piezoelectricity as described in (1), and external mechanical loading when applied. The model is implemented as a coupled thermo-mechanical model in Abaqus CAE by using user-defined subroutines (DISP, USDFLD and UMATHT) to assign thermal equivalent electrical properties to the membrane of each fibre, independently, based on the spike initiation [36], strain [2], [16] and voltage [31] generated at the same membrane, see (2) and (3). Electrical model parameters for other regions of the unmyelinated and myelinated fibres are taken from [33] and [2], respectively.

*E. Boundary Conditions*

In all cases, a voltage Gaussian distribution (with zero-mean and 0.4 variance) is the upper-threshold stimulation applied on Fibre #3 along its length, see Fig. 2, while the other fibres are activated only if the diffused charges from Fibre #3 generate an input voltage higher than the modulated threshold [36]. The voltage is applied at the nerve membrane layer through the DISP user-subroutine in Abaqus CAE, as a thermal load (i.e. equivalent to voltage [21]).The 3D distribution of charges on Fibre #3 modulates the activation of the other fibres, see Fig. 1. First, the electro-mechanical coupling has been assessed and validated by activating Fibre #3 with a pre-imposed strain condition (corresponding to a coupled left-shift (LS) of 0 or 20 $mV$ of the transmembrane voltage) assuming an intact or traumatized membrane, respectively, as in [2], [16].

An encastré boundary condition is enforced at the origin of each model, while no rotation nor shear is allowed. Then, two cases of mechanical loads applied at the bundle have been considered. As a first step to assess this novel coupling, only frequency-independent loading conditions are considered throughout, after the initial steady-state, so the solutions are quasi-static [21]. In the first case of applied mechanical loading, an instantaneous uniform compression is applied to the bundle to simulate injury conditions. Three values of pressure are modelled simulating mild (less than $55\ kPa$), moderate ($55 - 95\ kPa$) and severe (higher than $95\ kPa$) pressure [11]. In the second case, axial strain conditions that reproduce the uniaxial test in [10] are applied. Two values of instantaneous uniform stretch are applied as a displacement boundary condition to simulate 5 % and 14 % of total axial deformation, $\varepsilon$, at which the probability of inducing morphological injury during the elongation test is 5% and 25% respectively [10].

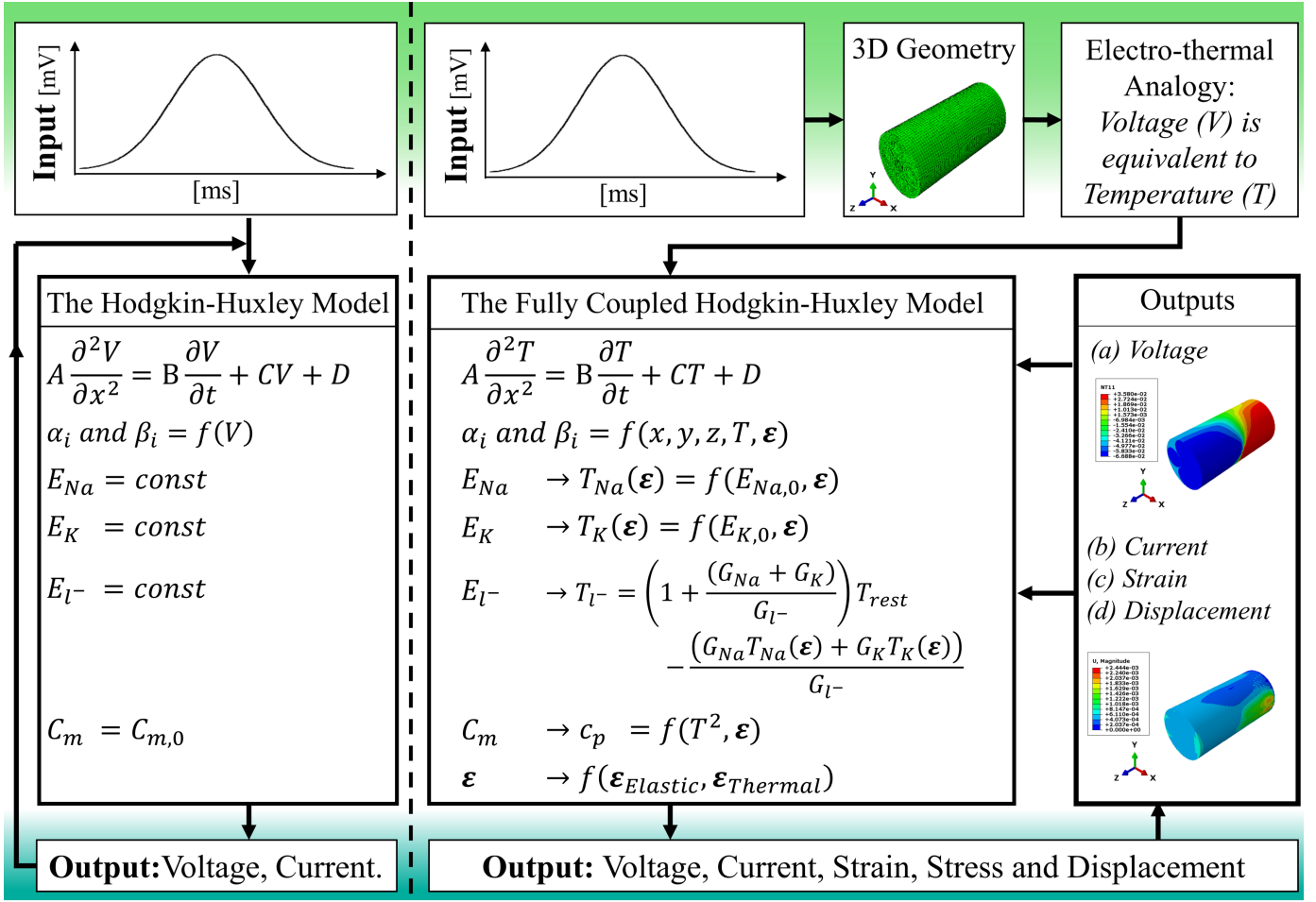

Fig. 2. Flowchart of the code describing the active behaviour of the nerve's membrane: on the left, the HH dynamics [24] and on the right, the fully coupled HH dynamics. Here, a Gaussian voltage distribution elicits the action potential in a 3D model of a nervous cell. By using electro-thermal equivalences, the HH model is implemented as an equivalent thermal process, in which the membrane's conductivity changes as in [24] and the capacitance, $C_m$, changes as in [31]. The HH parameters are changing based on the temperature (here the equivalent of voltage [21]) and strain at the membrane [2]. The strain $\varepsilon$ generated in the model is a function of temperature, $T$, and thermal expansion coefficients, see (3) and (4). Voltage, current, strain and stresses distribution are only a few of the 3D results released by Abaqus by equivalence.

*F. Validation*

The traumatic mechanically-induced damage [2], [16] is validated by using a coupled left-shift (*LS*) version of the HH model [16] reproducing experimental evidence [16], [17], as in [2]. The *LS* denomination refers to the induced shift of the transmembrane potential to positive values, simulating a nerve membrane subjected to mechanical trauma [2]. It is called coupled *LS* to highlight the close link of mechanical strain on the electrical properties of the nerve membrane. In contrast to [24], the ionic currents of the modified HH model include the fraction of affected ionic channels by the strain (*AC*) and coupled left-shift (*LS*) variables, see (5) and (6) [37]. When a trauma occurs at the nerve membrane, the ionic resting potentials undergo a voltage-shift ($LS$), here, equal to $20\ mV$, as a reference value for trauma-induced kinetic changes observed in experiments [37]. This allows to simulate uniaxial loading experiments [13], [38] of a nerve bundle, by implementing a pre-imposed shift in the HH dynamics, simulating the impact of applied mechanical strain at the nerve membrane. Then, the integrity of the nerve membrane is modelled as the fraction of the voltage-time dependent activation and inactivation variables (i.e. $m, n$ and $h$ [24]) of the HH model, shifted by $LS$ (here, indicated as $m_{LS}, n_{LS}$ and $h_{LS}$), see (5) and (6) [37]. The model accounts for the case where only a fraction of nodal channels ($AC$) undergoes a $LS$, while the rest of the membrane's channels, $(1 - AC)$, remain intact [16]. Here, only the extreme cases of the entire membrane being traumatized ($AC = 1$) or intact ($AC = 0$) are shown as illustrative examples. In (5), the sodium current, $I_{Na}$, and, in (6), the potassium current, $I_K$, [16]. The membrane potential is $V_m$, and the sodium, $\bar{g}_{Na}$, and potassium, $\bar{g}_K$, conductances are, respectively, equal to $120\ mS/cm^2$ and $36\ mS/cm^2$ [24].

$$I_{Na} = [m^3 h(1 - AC) + m_{LS}^3 h_{LS} AC](V_m - E_{Na})\bar{g}_{Na} \tag{5}$$

$$I_K = [n^4(1 - AC) + n_{LS}^3 AC](V_m - E_K)\bar{g}_K \tag{6}$$

This model is validated considering two cases of interest and two approaches. The first case (I) is the nerve membrane at physiological conditions, i.e. intact ($LS = 0$) and non-traumatized ($AC = 0$). In second case (case II), the membrane loses integrity ($AC = 1$) due to damage ($LS = 20\ mV$). Then, in the first approach, the reversal ionic potentials in (5)-(6), $E_{Na}$ and $E_K$, are constant values, equal to $50\ mV$ and $-77\ mV$ respectively [24], as in [16]. Here, the signals are changing because of the changes in conductance produced by changes in the voltage-dependent variables ($m_{LS}, n_{LS}$ and $h_{LS}$). In a second approach, the reversal potentials are changing with $LS$, see (4), as in [2]. Here, the signals are changing due to the changes in both the conductance and the reversal ionic potentials. Fig. 3 shows the changes in action potential due to the left-shift voltage implemented as in [16] (indicated with *) and as in [2] (indicated as **). Those two conditions are the ones that can be directly compared to [2], [16].

All the equations are collected in the Appendix in greater detail.

III. RESULTS

*A. Model Analysis*

The electrophysiological impairments associated with structural damage of the neuron mechanics affect the electrostriction and the electric field across the membrane, (3), while the piezoelectric properties of the nerve membrane are assumed to be constant, (2), [20], [25].

The results in Fig. 3 confirm the ability of the model to reproduce the membrane voltage and ionic current waveforms of a damage-induced injury. Here, the coupled $LS$ model of HH reproduces the changes in hyperpolarization in a damaged axon, left-shifting the affected ionic channels by $20\ mV$, as in slow-severe cases, as in [2], [16]. Fig. 3 refers to Fibre #3, the only fibre directly activated by the Gaussian voltage distribution input. Similarly, by diffusion, the membrane potential of the other fibres changes according to the level of damage and trauma on Fibre #3. Fig. 3 (a) shows the membrane potential and Fig. 3 (b) reports the ionic currents of the membrane (for leak ions, $I_l$-, potassium, $I_K$, and sodium, $I_{Na}$) vs. time in Cases I, II (*) and II (**). As shown, the mechanically-induced voltage-shift of the resting potentials leads to a time-delay and a lower amplitude of the voltage signal, due to leaky channels, similar to trends observed in [16] and [2]. In Case II (*), the action potential is shifted by $13.3\ ms$ and its peak at $16.9\ ms$ has a magnitude of about $10\ mV$. In Case II (**), the action potential is shifted by $13.7\ ms$ and the maximum magnitude is about $-46.5\ mV$ at $25.1\ ms$.

To illustrate the utility of the model in predicting the nerve mechanical response, Figs. 4 (a) and 4 (b) show the corresponding displacement vs. time of an unmyelinated nerve membrane and myelinated nerve membrane, respectively, in a damaged bundle along the radial direction (i.e. the $x - axis$). Data are taken at the position of maximum radial displacement on Fibre #3. Fig. 5 shows the total displacement distribution in an unmyelinated bundle (from (a) to (d)) and in a myelinated bundle (from (e) to (h)) with $LS = 20\ mV$ and $AC = 1$. In both cases, data are plotted at the time instant when the mechanical displacement is at its maximum.

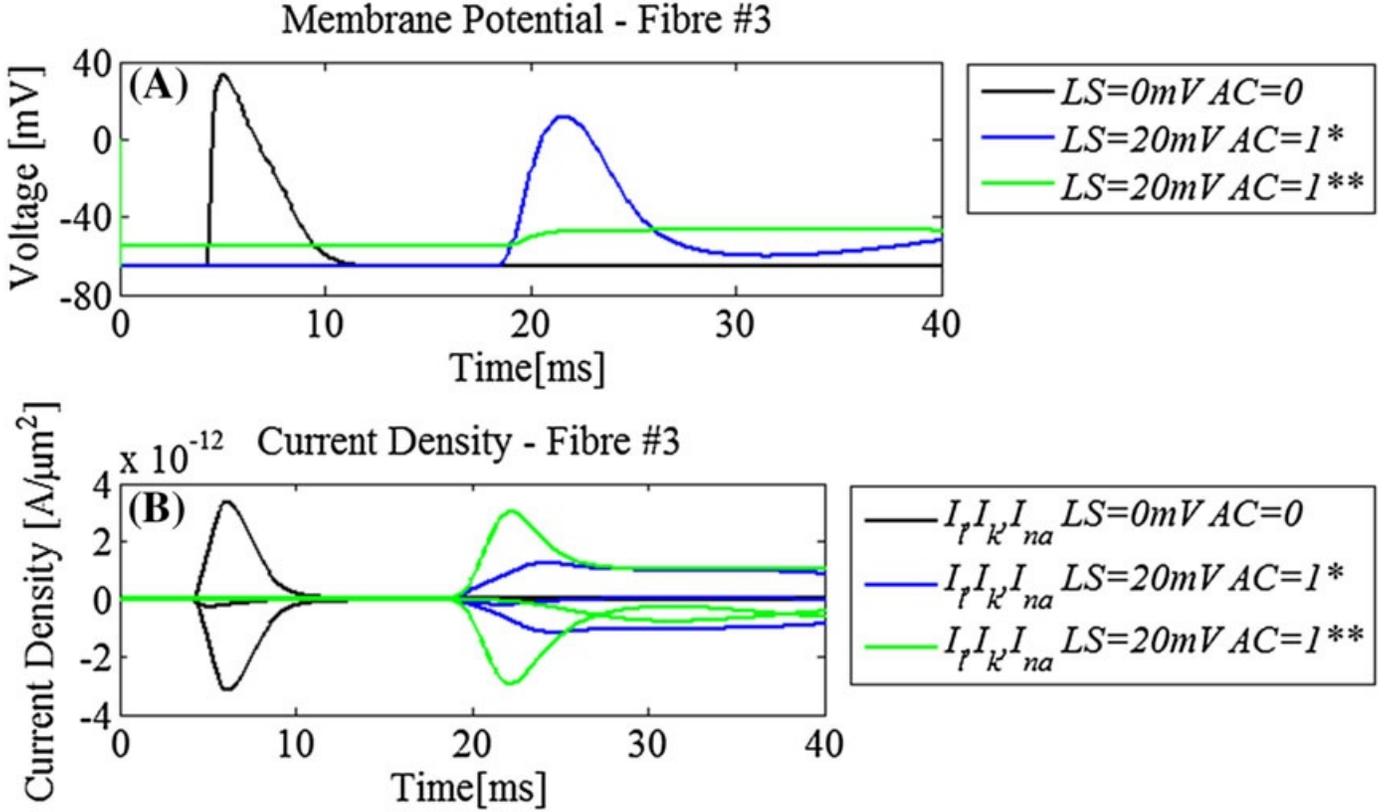

Fig. 3. (a) the membrane's potential in Case I, action potential ($LS = 0mV; AC = 0$); Case II, damaged traumatized membrane ($LS = 20mV; AC = 1$). In *, the reversal ionic potentials have constant values as in [16], and, in **, they change according to equation (4), as in [2]. In (b), the Current Density $[A/\mu m^2]$ on Fibre #3 for the cases considered.

For each case in Fig. 3, residual compressive forces are found in unmyelinated and myelinated fibres, see Fig. 4 and 5, due to the biophysical activity of the nerve membrane [20], [31]. In an unmyelinated bundle, the displacement peak is $-3.3\ nm$ for Case I, $-2.5\ nm$ for Case II (*), and $-0.6\ nm$ for Cases II (**). In a myelinated bundle, the peak is $-0.81\ nm$ for Case II (*) and $-0.27\ nm$ for Case II (**). The peaks of the radial displacements occur at the membrane potential peaks for each case. In a myelinated bundle, lower displacements are found at the membrane than in the unmyelinated case, see Fig. 4 (b) and 5 (e)-(h). The myelin layer constrains the deformation of the nerve membrane, reducing the displacement by about an order of magnitude on the same unmyelinated fibre, see Fig. 4 (b) and Fig. 5 (h). Assuming the Ranvier node regions are aligned in a bundle, the total displacement of the bundle is driven by the same deformation as the activated fibre, see Fig. 5 (h).

B. *Mechanical Loading Cases of Interest*
In this section, the fully coupled electro-mechanical model of Fig. 2 is applied to investigate different mechanical loading conditions directly using equations (1-4), instead of as a simulated voltage-shift.

1. *Pressure Loads*
Fig. 6 (a) and Fig. 6 (b) show the hyperpolarization and current densities of an unmyelinated bundle and a myelinated bundle under mild ($25kPa$), moderate ($68kPa$) and severe ($192kPa$) pressures inducing Traumatic Axonal Injuries (TAI) [11]. The case of extreme pressure ($1GPa$) is also considered. In contrast to the reference case of an intact nervous cell ($P = 0\ kPa$ and $AC = 0$), all pressure loads reduce the magnitude of the action potential and shift it over time, correlating with results in Fig. 3. Here, the strain applied at the nerve membrane by compressing the bundle shifts the ionic resting potentials of the fully coupled HH model by a quantity which varies depending on the magnitude of the applied load, see (4). Thus, only the $AC$ variable is, here, considered.

Compression levels in the range of $25\ kPa$ to $192\ kPa$ have a similar impact on the signal transmission both in terms of reduced magnitude and shift over time; this is found to be due to similar strain values read at the nerve membrane, because of the linear elastic assumption. In these cases, the peak of the action potential is about $-17\ mV$ with $16\ ms$ of time-delay. Here, only slight differences are found for a traumatized vs. non-traumatized nerve membrane ($AC = 1$ vs. $0$) when mild-to-severe pressures are applied, and therefore only results for $AC = 0$ are included, see Fig. 6.

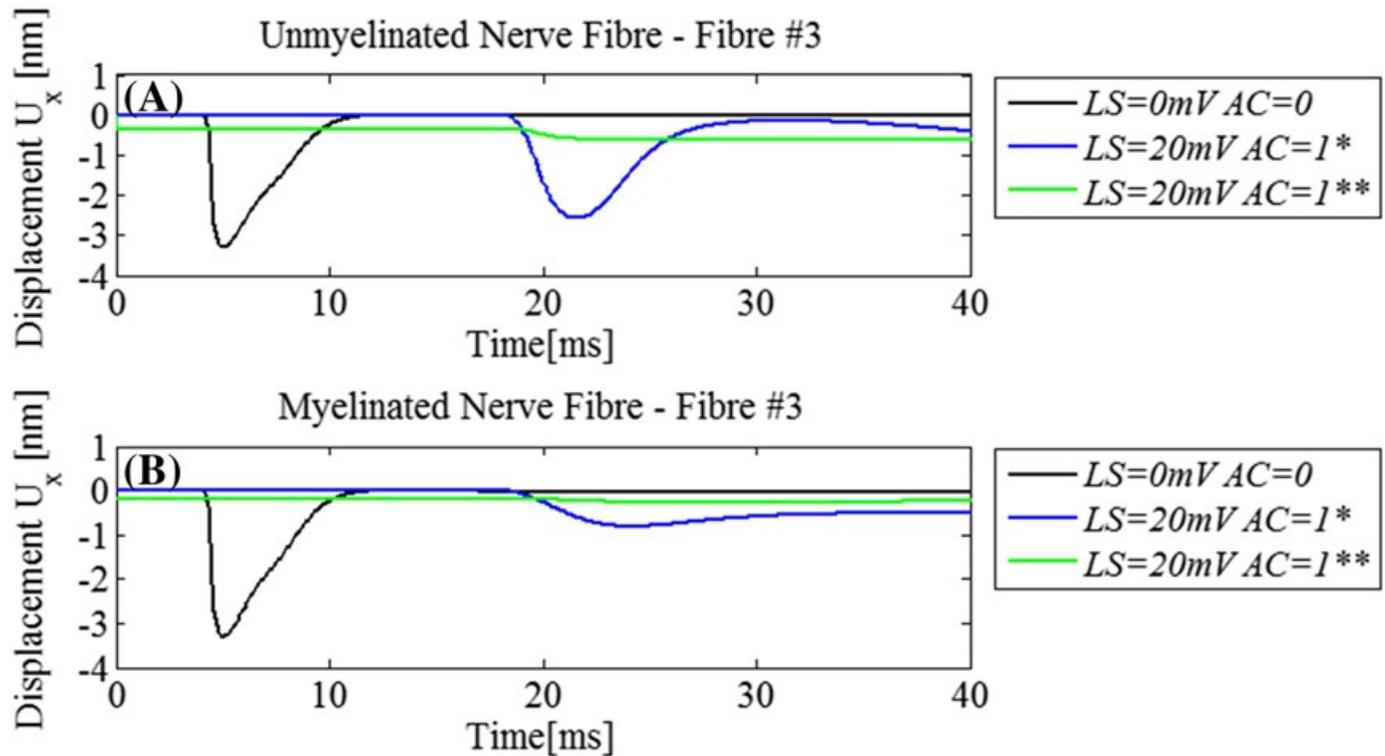

Fig. 4. (a): mechanical displacement of the unmyelinated nerve membrane; (b): displacement of a myelinated nerve membrane of the Fibre #3 in the four cases considered (see text) along the radial direction in the bundle (i.e. $x - axis$), $U_x$.

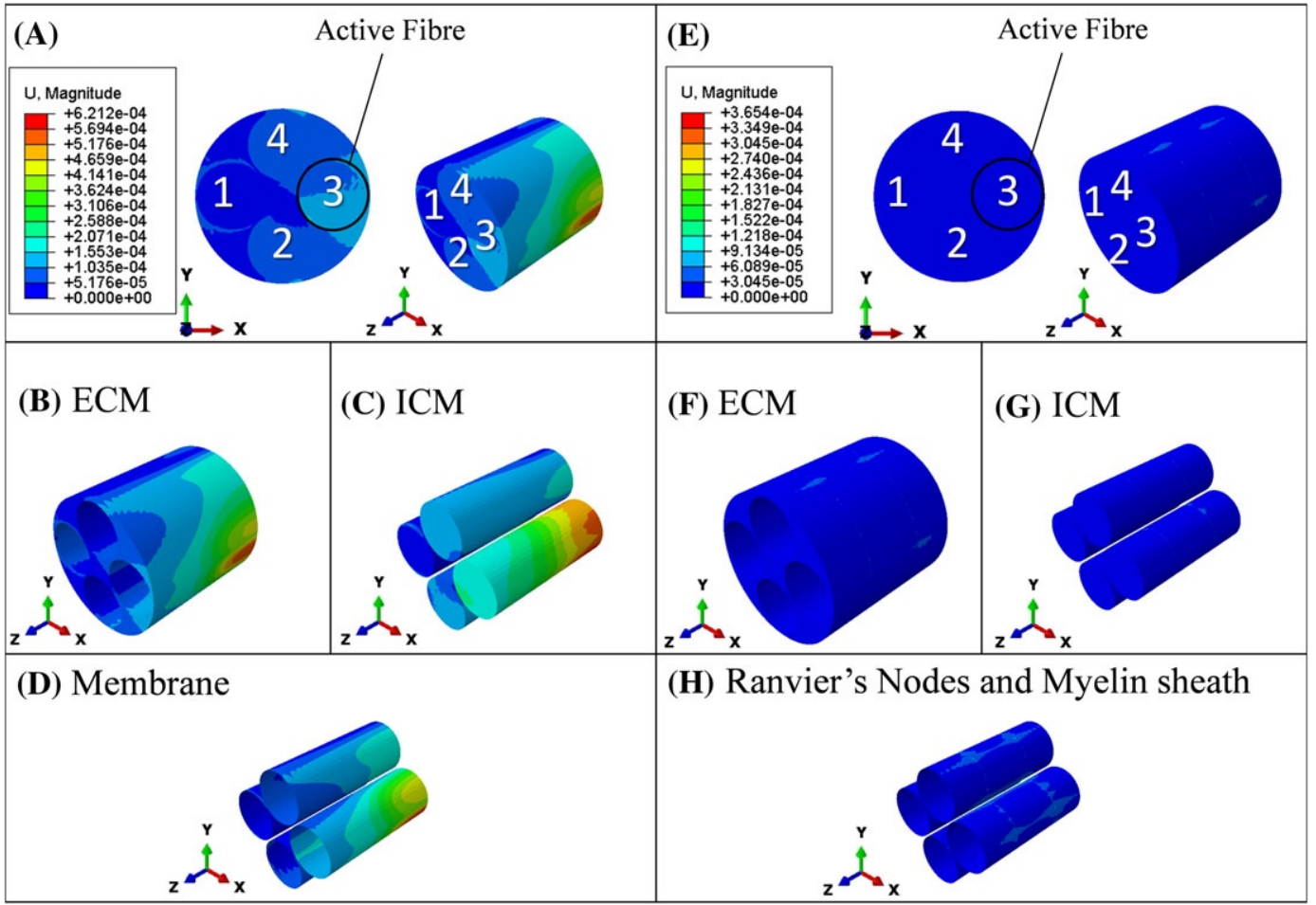

Fig. 5. Frontal and isometric views of the total displacement in (a) an unmyelinated and (e) a myelinated nerve bundle model. In (b) and (f), the ECM; in (c) and (g), the ICM of the two bundle types. In (d), the isometric view of the nerve membrane and, in (h), the Ranvier nodes and the myelin sheath of the myelinated bundle. (a)-(d) and (e)-(h) are Case II (**) [2] applied to an unmyelinated and myelinated nerve bundle. Data are taken at the peak of the action potential in both cases.

In contrast, application of extreme pressure changes the resting potential of ions reducing the current flow across the nerve membrane, see Fig. 6 (b), and reducing the magnitude of the action potential to zero, see Fig. 6 (a). An extreme pressure leads to both a reduction in magnitude and an increase of the voltage baseline up to $-24\ mV$ for an intact membrane, and up to $-7mV$ for a traumatized membrane, see Fig. 6 (a) and (b). Here, the homeostatic balance of charges across the nerve membrane vanishes with an extreme pressure applied over a traumatized nerve membrane ($AC = 1$), because the strain levels at the nerve membrane are close to the threshold value assumed in (4).

With mild-to-severe pressure loads from $25\ kPa$ to $192\ kPa$, the contraction of the nerve membrane due to the electrostriction has a similar trend to the case of $LS = 0\ mV$ and $AC = 1$, see Fig. 3. Figs. 7 (a) and 7 (b) show the radial mechanical displacement on Fibre #3 of an unmyelinated and myelinated fibre, respectively. In both unmyelinated and myelinated bundles, the action potential is shifted through time durations by about $4.70\ msec$ for the extreme pressure case (not shown here) and by $16.2\ msec$ for the mild-to-severe cases [11], respectively, see Figs. 7 (a) and 7 (b). Higher deformation at the nerve membrane changes the ionic resting voltages leading to higher mechanical displacements, see (4) and Fig. 8.

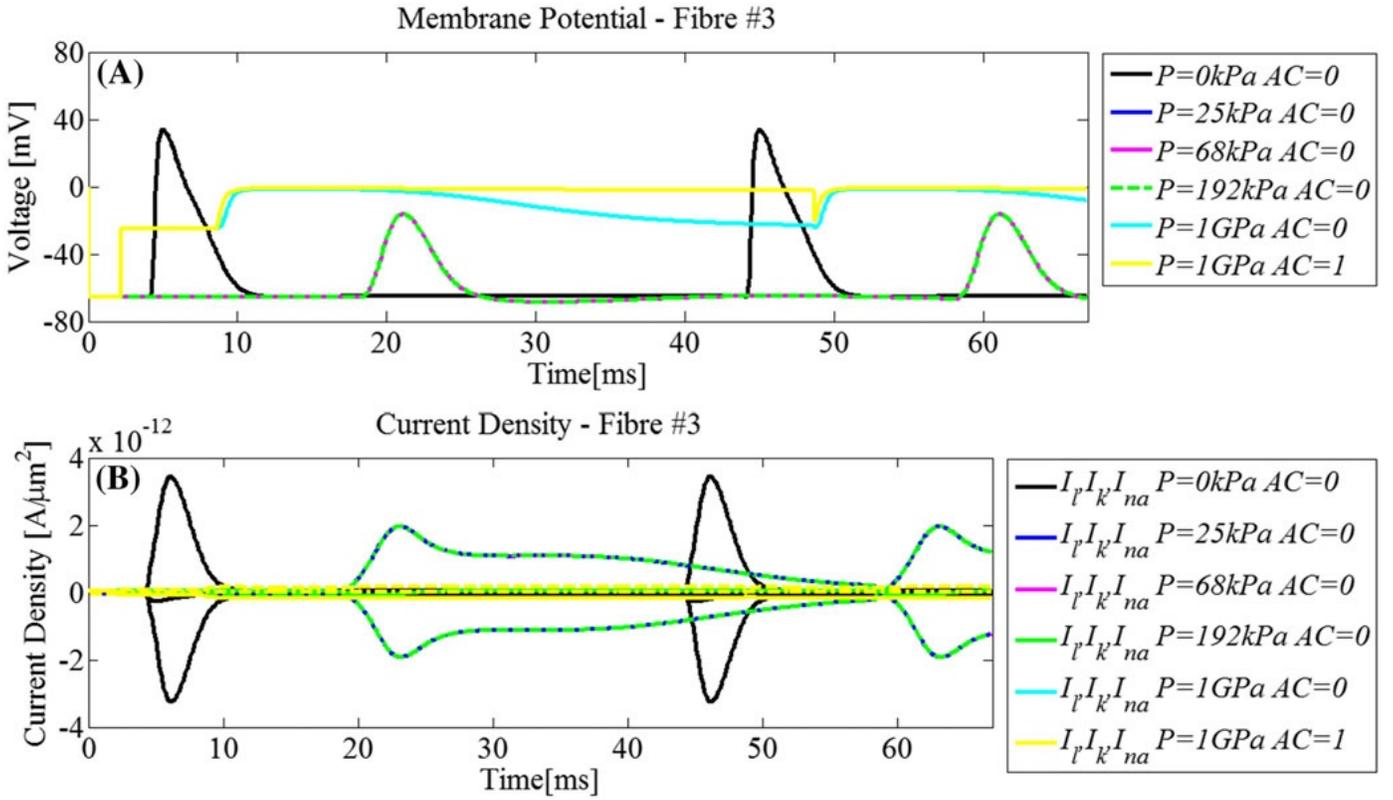

Fig. 6. (a) Membrane Potential [$mV$] and (b) Current Density [$A/\mu m^2$] on Fibre #3 in the unmyelinated bundle under mild ($25 kPa$), moderate ($68 kPa$) and severe ($192 kPa$) pressures [11]. The extreme case of $P = 1 GPa$ is also considered. AC is the fraction of affected ionic channels by the strain: $AC = 0$ is for an intact membrane and $AC = 1$ for a traumatized membrane [16]. Data are the maximum radial displacement of a node on Fibre #3.

Then, for the mild to severe cases of TAI-induced pressure [11], the resting voltage potentials are changed due to the induced deformation in the bundle and the magnitude of the action potential is, hence, reduced [2], see Fig. 7. The peak of the action potential is higher in a compressed unmyelinated bundle than in a compressed myelinated one.

Within the range of pressure levels considered, an unmyelinated layer displaces according to the charges exchanged across the nerve membrane, see Fig. 7. On Fibre #3, for the case of mild, moderate and severe pressures, the peak is $-1\ nm$, $-1.10\ nm$, and $-1.40\ nm$, respectively, see Fig. 7 (a). In a myelinated bundle, the charge-induced displacement of a myelinated fibre is much less than in an unmyelinated bundle and therefore its displacement is more in response to the loading condition than to electrostriction, see Fig. 7 (b). Here, on Fibre #3, for the cases of mild, moderate, and severe pressures, the peak is $-0.14\ nm$, $-0.40\ nm$, and $-1.20\ nm$ respectively, see Fig. 7 (b).

This model shows greater membrane displacements in an unmyelinated fibre than in a myelinated fibre, see Fig. 7 (a) and 7 (b). The myelin layer constrains the deformation of the Ranvier nodes, which are the only regions throughout the fibre to show voltage-induce membrane displacement [20]. At the nodes, the applied compression acts in opposition to the electrostriction, because of the negative value of the action potential.

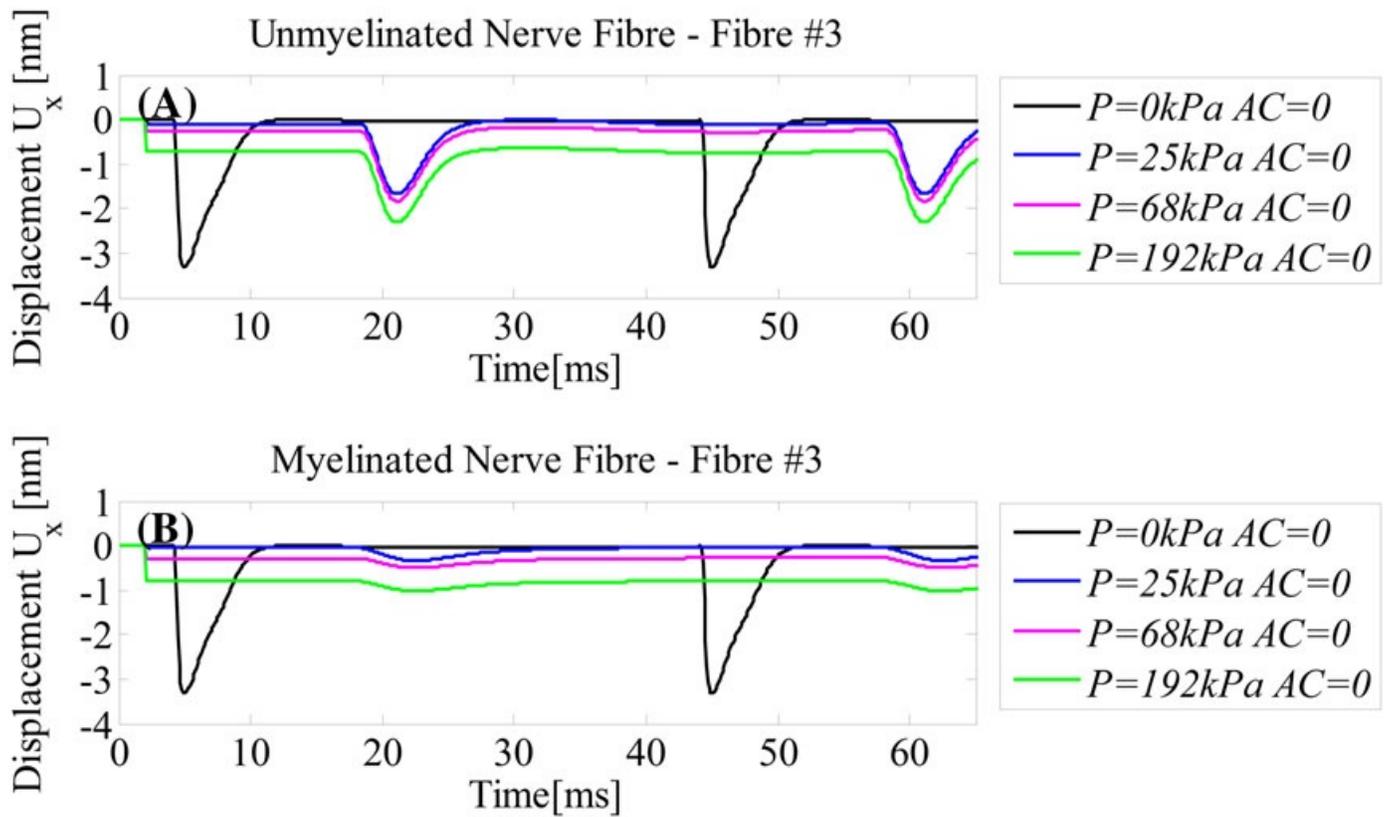

Fig. 7. Radial displacement [$nm$] of (a) an unmyelinated bundle and (b) a myelinated bundle. Uniform applied pressures are classified as mild ($25kPa$), moderate ($68kPa$) and severe ($192kPa$) pressures [11]. Data are the maximum radial displacement of a node on Fibre #3 in both cases.

2. *Displacement Loads*

In Figs. 8 (a)-(b) and 8 (c)-(d) a displacement boundary condition applied along the length of the unmyelinated and myelinated bundles, respectively, simulates 5% and 14% of total deformation [7].

In an unmyelinated bundle, see Fig. 8 (a)-(b) and Fig. 9 (a)-(d), the action potential is shifted through time durations by about $19\ ms$ for $\varepsilon = 5\%$ and $AC = 0$, and by $21\ ms$ for $\varepsilon = 5\%$ and $AC = 1$. For higher deformation, the action potential is delayed by $8\ ms$, showing similar results both with $AC = 0$ and $AC = 1$, representing the loss of ionic gradient across the nerve membrane, see Fig. 8 (b).

In contrast, in a myelinated bundle, see Fig. 8 (b)-(d) and 9 (e)-(h), no significant differences have been found between intact and traumatized membranes. Here, for 5% and 14% deformation, the membrane potential is shifted up to $-60mV$ and $-43mV$, respectively, while the peak of the action potential is shifted at $23\ ms$ and $17\ ms$, respectively, see Fig. 8 (c). Lower current density at the Ranvier nodes is mainly due to lower voltage gradient and higher localized strain, see Figs. 8 (d) and 9 (h).

## IV. Discussion

In contrast to previous studies [1], [2], this paper shows the advantages of a fully coupled electro-mechanical 3D framework to investigate the details of neural activity, combining real-time fully coupled electro-mechanical phenomena, modulated threshold for spiking activation, and independent alteration of the electrical properties for each fibre in the 3-layer nerve bundle, made of membrane (or Ranvier's nodes and myelin sheath), ICM and ECM. The electro-mechanical coupling, based on electro-thermal equivalences [22], [23], [26], allows for reliable simulation of changes in electrostriction and neural activity due to mechanical damage, as seen in experiments [10], [11], [16].

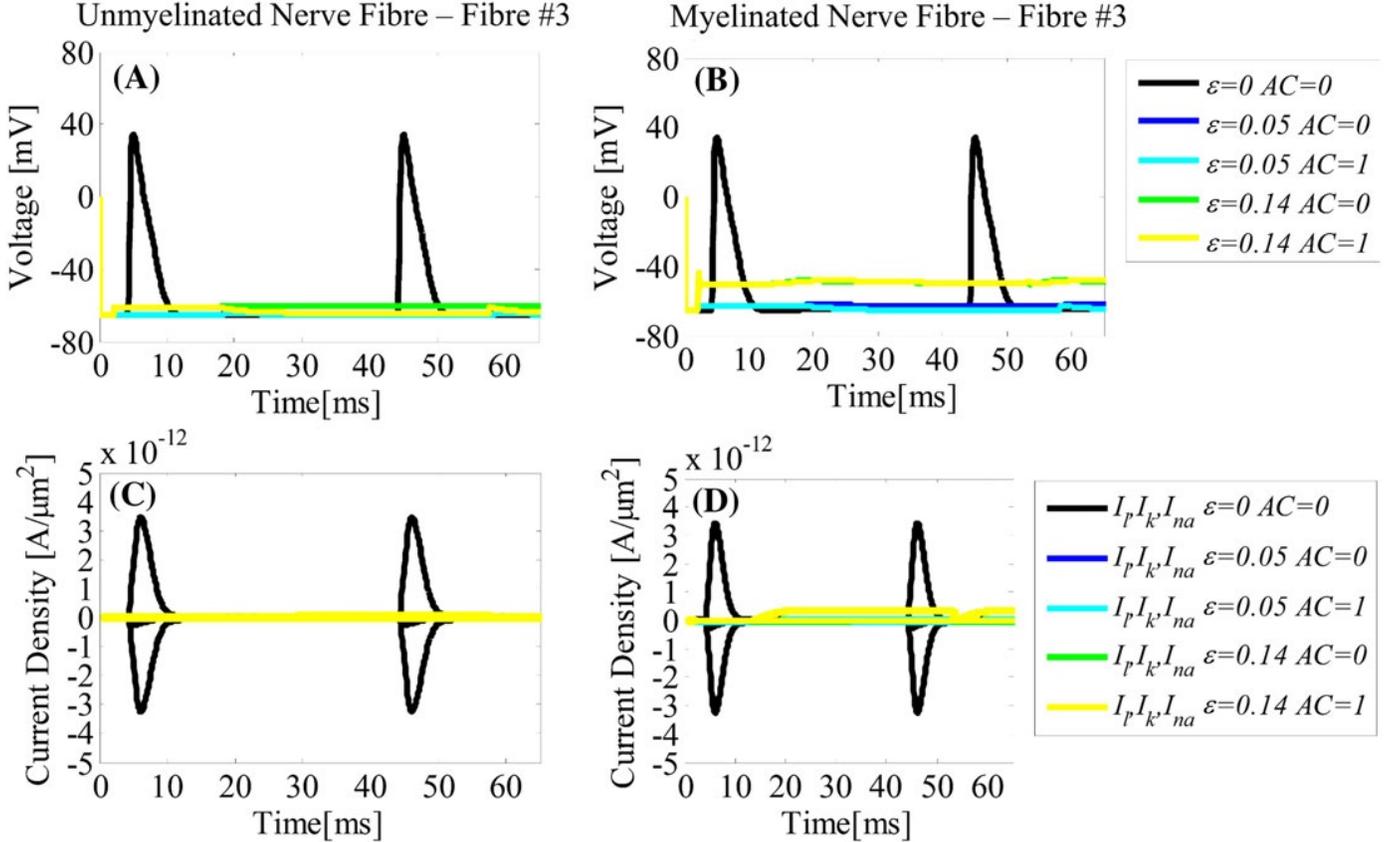

Fig. 8. (a)-(b) Membrane Potential $[mV]$ and (c)-(d) Current Density $[A/\mu m^2]$ on Fibre #3 in the unmyelinated bundle and myelinated bundle, respectively, under 5% and 14% of total deformation $\varepsilon$ applied [10]. AC is the fraction of affected ionic channels by the strain: $AC = 0$ is for an intact membrane and $AC = 1$ for a traumatized membrane [16]. Data are taken at the maximum displacement along the bundle middle axis, i.e. $z-axis$, on Fibre #3.

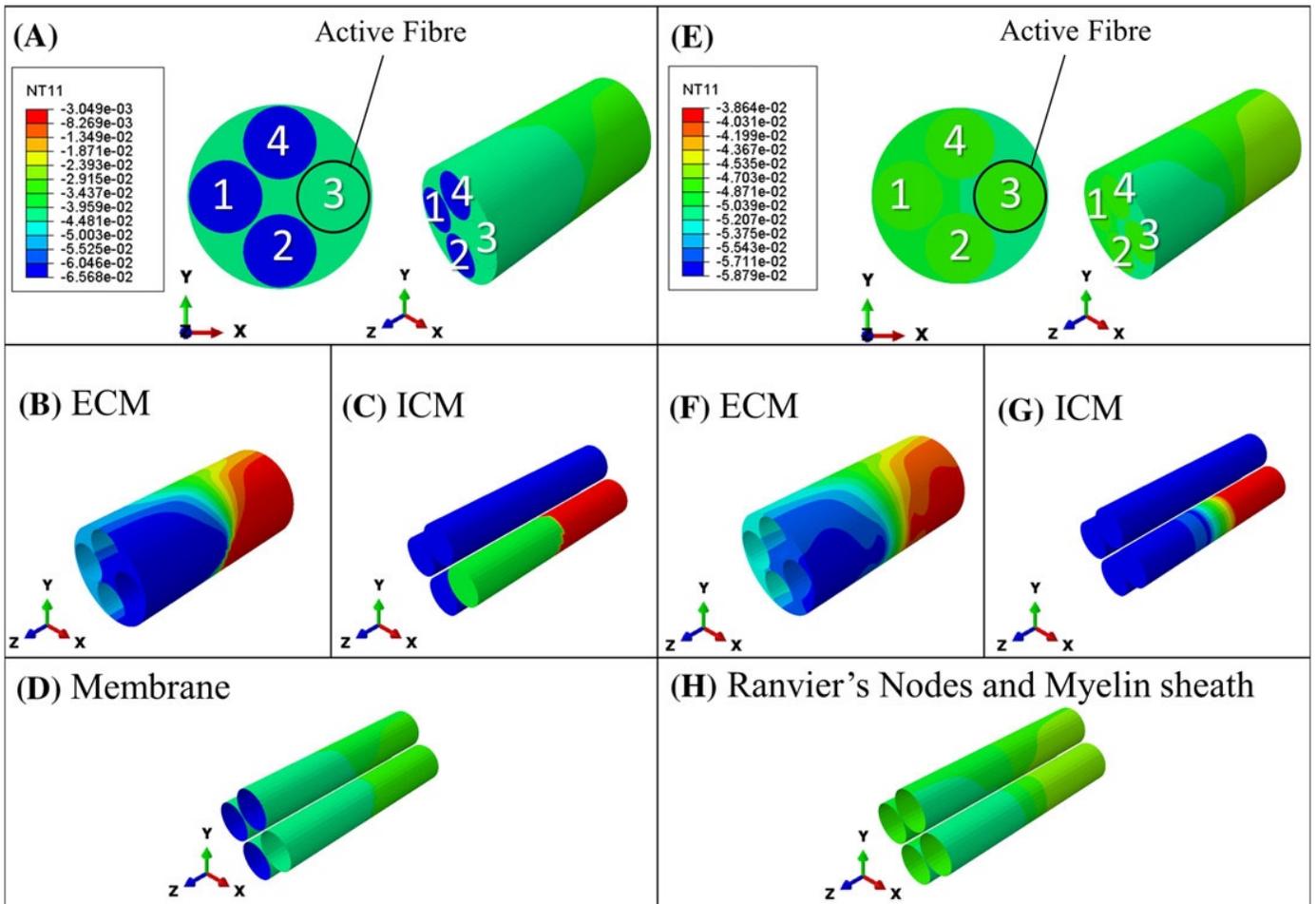

Fig. 9. Voltage distribution ($NT11$) in an unmyelinated nerve bundle, in (a)-(d), and in a myelinated nerve bundle, in (e)-(h), for 5% elongation. Frontal and isometric view of an unmyelinated and myelinated nerve bundle model in (a) and (b), respectively. In (b) and (f), the ECM, in (c) and (g) the ICM of the two bundle types. In (d), the isometric view of the nerve membrane and, in (h), the Ranvier's nodes and the myelin sheath of the myelinated bundle.

In this study, two cases of interest provide insights into the electrophysiological impairments of axonal injury due to sudden TAI-induced pressures and displacements. Differences in signal transition arise in the bundle for each fibre, depending on the fibre type. In the bundle, Fibre #3 is activated by imposing a voltage Gaussian distribution, while the other fibres are activated based on the voltage gradient from the active fibre and total strains (i.e. elastic and thermal strain) read at the nerve membrane. Here, the nerve membrane integrity depends on the ionic resting potentials which are a function of the voltage and total strain at the nerve membrane. If the strains at the nerve membrane are lower than 0.21 [10], the resting voltage potentials changes accordingly to the strain intensity. Alternately, no ionic gradient can be simulated at the membrane because the resting voltage values are zero [2], as in the case of pressure higher than $192 kPa$ and elongation greater than 14%, see Figs. 7 and 9. Unmyelinated nerve fibres show the greatest changes with mechanical loads due to the higher current density per area exchanged at the membrane. In contrast, lower density of ions per area can flow thought the membrane at the Ranvier nodes, while the myelin layer constrains the electrostriction accompanying the neural activity.

Results show that in the myelinated bundle the activation of the fibres is compromised because the fibre is not able to follow the pattern of activation under pressure and displacement loads. These two loading conditions are cases of clinical interest. A constant pressure on an axon is representative of injuries [1], [3] and plaques in demyelinating diseases, acting as conduction block for action potentials of neighbouring axons inhibiting the myelinisation process. Instead, elongation tests might explain the relatively long time period needed for spontaneous membrane repair after fast strain [13].

Finally, as highlighted earlier, the results refer to a cylindrical bundle made of four identical fibres with characteristics within the range of the human optic axon [32]. Although real nerve bundles are made of a higher number of fibres with different calibre [32], the use of a simplified geometry was needed to assess electro-mechanical equivalences in a 3D FE model and to limit the computational cost. On-going work is focused on extending this technique to nerve bundle with different calibre (made of unmyelinated, myelinated and mixed fibres) and multiple activation of fibres.

## V. CONCLUSION

We propose a fully coupled electro-mechanical framework for modelling the biophysical phenomena accompanying neural activity. Here, the coupling is based on an electro-thermal analogy by modelling the piezoelectric effect as a thermal expansion phenomenon in Abaqus CAE 6.13-3. This approach allows us to model and generate insights into aspects of neural activity, such as electrostriction and piezoelectricity, and to correlate these with experimental observations. The model, built on previously published work [22], [23], [26], generates a fully 3D simulation of ion channel leaking for nerve fibres under pressure and displacement loads. In conclusion, in this model:
- Time-shift, signal magnitude and nerve membrane potential baseline are dependent on the total strain, voltage and size of the fibre;
- Lower strain and lower electrophysiological changes are found in myelinated fibre Vs. unmyelinated fibre;
- Unmyelinated fibre are needed to share information across the fibre, rather than throughout its length.

This model can contribute to the understanding the causes and consequences of TBI and DAI to improve diagnosis, clinical treatments and prognosis by simulating the mechanical changes accompanying the changes in signal transmission in TAI-induced loading conditions.

## VI. ACKNOWLEDGMENT

The authors gratefully acknowledge funding from the Galway University Foundation, the Biomechanics Research Centre, and the Power Electronics Research Centre, College of Engineering and Informatics, NUI Galway, Ireland.

## APPENDIX

The fully coupled Hodgkin and Huxley model reported in this work is based on the electro-thermal equivalences [39] and on the electro-mechanical coupling, see (A.3) and (A.4).

First, (A.1) and (A.2) show the implementation of the 3D Heat Diffusion Equation as Cable Equation, see [39]. The electrical material properties are substituted to the corresponding thermal properties, as discussed in [39]. In the Heat Equation (A.1), $T$ is the temperature [$°C$ or $K$], $\rho$ is the mass density [$kg/m^3$], $c_p$ is the specific heat capacity [$J/(kg\ K)$], $k$ is the thermal conductivity [$W/(mK)$], and $Q$ is the heat source density [$W/m^3$]; correspondly, in the Cable Equation (A.2) applied to the nerve, $V_m$ is the transmembrane potential [$V$], $S_v$ [$1/m$] denotes the surface-volume ratio, $C_m$ [$F/m^2$] is the nerve membrane capacitance, $\sigma$ [$S/m$] is the electrical conductivity of the membrane, and $I_{ionic}$ [$A/m$] is the ionic current.

$$\rho c_p\ \partial T/\partial t - \nabla \cdot (\boldsymbol{k}\nabla T) + Q = 0 \qquad (A.1)$$

$$C_m S_v\ \partial V_m/\partial t - \nabla \cdot (\sigma_e \nabla V_m) + S_v I_{ionic} = 0 \qquad (A.2)$$

Second, (A.3) and (A.4) display the piezo-elastic strain-stress relationship vs. the thermo-elastic strain-stress relationship, respectively (see Method – Section A). In the electro-mechanical coupling, the piezoelectric strains are simulated by using the analogue quantity corresponding to the piezoelectric coefficients, i.e. the expansion coefficients $\boldsymbol{\alpha}$.

$$\varepsilon = \beta\sigma + \delta(\Delta V/h) \tag{A.3}$$

$$\varepsilon = \beta\sigma + \alpha\Delta T \tag{A.4}$$

Then, simulating quasi-static solutions [39], (A.5) shows the balance of ionic currents modelling the nerve membrane, as in the Hodgkin and Huxley model [24].

$$\begin{aligned} C_M \frac{\partial V}{\partial t} &= \bar{g}_{Na}m^3h(V - E_{Na}) + \bar{g}_K n^4(V - E_K) + \bar{g}_l(V - E_{l^-}) \\ &= G_{Na}(V - E_{Na}) + G_K(V - E_K) + G_{l^-}(V - E_{l^-}) = I_{Na} + I_K + I_{l^-} \end{aligned} \tag{A.5}$$

Following trauma, electrical alterations can be due to: (i) the loss of integrity of the membrane [16] where a fraction of affected ionic channels ($AC$) is assumed to be traumatized; and (ii) transmembrane voltage shift ($LS$) to positive values [16], representing a different osmotic gradient due to the trauma-induced strains. Sodium, $I_{Na}$, potassium, $I_K$, and leak ions current, $I_{l^-}$, are shown in (A.6), (A.7) and (A.8), respectively (see Method – Section F).

$$I_{Na} = [m^3h(1 - AC) + m_{LS}^3 h_{LS} AC](V_m - E_{Na})\bar{g}_{Na} \tag{A.6}$$

$$I_K = [n^4(1 - AC) + n_{LS}^3 AC](V_m - E_K)\bar{g}_K \tag{A.7}$$

$$I_{l^-} = \bar{g}_{l^-}(V_m - E_{l^-}) \tag{A.8}$$

Additionally, the reversal potentials of potassium, (A.9), sodium, (A.10), and leak ions, (A.11), as the axonal component of the total strain applied at the bundle is changing (see Method – Section C). The membrane resting potential is $V_{rest}$.

$$E_{Na}(\varepsilon_m) = E_{Na0}(1 - (\varepsilon_m/\tilde{\varepsilon})^\gamma) \tag{A.9}$$

$$E_K(\varepsilon_m) = E_{K0}(1 - (\varepsilon_m/\tilde{\varepsilon})^\gamma) \tag{A.10}$$

$$E_{l^-} = \left(1 + \frac{G_{Na} + G_K}{G_{l^-}}\right)V_{rest} - \frac{G_{Na}E_{Na}(\varepsilon_m) + G_K E_K(\varepsilon_m)}{G_{l^-}} \tag{A.11}$$

Finally, the electrical capacitance per unit area is changing with voltage as in (A.12) (see Method – Section A).

$$C(V) = C(0)[1 + \vartheta(V + \Delta\varphi)^2] \tag{A.12}$$